\documentstyle[twocolumn,prl,aps]{revtex}
\tighten
\begin{document}
\title{Generic morphologies of viscoelastic dewetting fronts}
\author{Stephan Herminghaus, Ralf Seemann, Karin Jacobs}
\address{Applied Physics Lab., University of Ulm, D-89069 Ulm,
Germany}
\date{\today}
\maketitle

\begin{abstract}
A simple model is put forward which accounts for the occurrence of
certain generic dewetting morphologies in thin liquid coatings. It
demonstrates that by taking into account the elastic properties of
the coating, a morphological phase diagram may be derived which
describes the observed structures of dewetting fronts. It is
demonstrated that dewetting morphologies may also serve to
determine nanoscale rheological properties of liquids.
\end{abstract}


When a thin liquid film beads off a solid substrate, it is
eventually transformed into an ensemble of individual liquid
droplets, the arrangement of which may vary strongly according to
the basic mechanisms involved in the dewetting process
\cite{Srolovitz1,Redon,Reiter,Karinsdiss,Lambooy,Stange,Science}.
In by far most cases, dewetting is initiated by heterogeneous
nucleation of individual holes in the initial film, thus forming
contact lines between the film surface and the substrate. The
surface forces acting upon these contact lines give rise to moving
dewetting fronts, the motion, shape, and interplay of which
largely determine the final dewetting morphology.

It would be very desirable to have a theory describing the
dynamics of these fronts precisely enough for extracting
information on the dewetting mechanisms from the characteristics
of their shape and the final droplet structure. However, the task
of setting up such a theory has proved untractable so far. Solving
the Navier-Stokes equation with the moving contact line as a
boundary condition is particularly complicated and entails {\it ad
hoc} assumptions on the dynamics at the contact line
\cite{Moriarty,Greenspan,Troian}. Furthermore, it has meanwhile
become clear that viscoelastic effects, which are not described by
the Navier-Stokes equation, may be decisive for the evolving
morphology in both advancing and receding fronts
\cite{Karinsdiss,Spaid,PhysicsOfFluids,falaises,Unterschwinger}.
In particuilar, it has recently been shown
\cite{falaises,Unterschwinger} that `elastic´ dewetting fronts may
strongly differ, as to their shape and dynamics, from Newtonian
ones \cite{LangmuirGrowth}, and it has been pointed out that these
findings are not accounted for by theory so far \cite{falaises}.
Consequently, what is needed is a handy theoretical model, which
is just as detailed as necessary to describe the main physics
involved in the dewetting dynamics of viscoelastic films. It is
the purpose of this paper to propose such a model.

Let us first recall the main features to be explained, by
summarizing the generic front morphologies observed so far.
Fig.~\ref{Observations} shows dewetting fronts in liquid
polystyrene films beading off silicon substrates (the fronts move
from left to right). The profiles are obtained by scanning  force
microscopy from the rims of circular dry patches nucleated in the
films \cite{GuenterMacromolecules,LangmuirRupture,Nucleation}.
Most commonly, one observes profiles as those shown in
fig.~\ref{Observations}a, with a simple decay from the crest into
the undisturbed film to the right. However, when the molecular
weight of the polymer is small enough \cite{Unterschwinger}, the
shape may be qualitatively different. This is shown in
fig.~\ref{Observations}b, which has been obtained with polystyrene
films with a molecular weight of only 2 kg/mol. A ditch is clearly
visible in front of the crest, and even another small elevation to
the right of the ditch is present as well, such that the front
appears as a damped oscillation. The only difference between the
film materials used for fig.~\ref{Observations}a and
\ref{Observations}b is their molecular weight, and thus their
viscoelastic properties. If the film is not too thick, the ditch
may reach the substrate and pinch off the crest from the film,
forming a new contact line. This happens repeatedly,  such that a
series of isolated crests forms in a cascade of pinch-off events,
as shown in fig.~\ref{Observations}c.

We shall now create a suitable mathematical description of these
findings. An incompressible viscoelastic fluid may be described by
the force balance \cite{LL1,LL2,Questions}
\begin{equation}
\eta \Delta j = \nabla p - E \Delta \phi
\label{ForceBalance}
\end{equation}
where $\eta$ is the viscosity, $j$ is the material current in the
film, $p$ is the pressure, $E$ is Young's modulus, and $\phi$ is a
local displacement field describing the strain
\cite{Questions,SurfaceMelting}. In the present treatment, we will
adopt the lubrication approximation \cite{BrochardExtrapLength},
thereby neglecting the normal ($z$-) components of the current and
the pressure gradient, and consider quantities averaged over the
film thickness, $h$. We then can rewrite eq.~(\ref{ForceBalance})
as $J(x) = -C \partial_x (p-\alpha E\partial_x\phi)$, where $J$ is
the total current in the film, and $\alpha$ is a numerical factor
which characterizes the flow profile. The latter depends, e.g., on
the friction of the film at the substrate, and it is clear that in
the case of full slippage, the coupling of the flow to the strain
will not be the same as in the absence of slip. These effects are
difficult to treat explicitly, and thus will be absorbed here into
$\alpha$. Similarly, $C$ depends upon the viscosity of the film,
$\eta$, and its friction coefficient at the substrate surface,
$\kappa$. If the latter is infinite (Poiseuille flow with no slip
at the substrate), we have $C=h^3/3\eta$, while for $\kappa
\rightarrow 0$ (plug flow), $C=h^2/\kappa$
\cite{LangmuirGrowth,BrochardExtrapLength}.

Since the derivative of the displacement is the strain,
$\sigma(x,t)=\partial_x\phi$, and excess pressure in the film comes about from
the curvature of its surface, we have
\begin{equation}
J(x,t) = C \partial_x \{ \gamma\partial_{xx}\zeta(x,t) + \alpha
E\sigma(x,t) \}
\end{equation}
where $\gamma$ is the surface tension, and $\zeta(x,t)$ is the
vertical displacement of the film surface. We explicitely neglect
long range wetting forces acting through the film, such as van der
Waals forces. In sufficiently thick films, which dewet by
nucleation \cite {LangmuirRupture,Reconciliation}(such as assumed
above), these do not play a significant role.

Since the strain in the film may decay by internal relaxations of
the material according to $\partial_t \sigma = -\omega_0 \sigma$,
its coupling to the material current can be written as
\cite{Questions,SurfaceMelting}
\begin{equation}
(\partial_t+\omega_0) \sigma = \partial_x J/h
\end{equation}
where we have neglected the nonlinear convective term $J\partial_x
\phi$. This restricts our discussion to small excursions $\zeta$,
but enables a linear treatment. By combination of the above
equations with the continuity equation, $\partial_t \zeta +
\partial_x J = 0 $, it is easy to obtain the equation of motion
for $\sigma$ and $\zeta$. A precondition for elasticity to play a
significant role is that $\omega_0 << \omega$, in which case the
dispersion relation reads
\begin{equation}
\omega = \omega_0 + C q^2 \left(\frac{\alpha E}{h} + \gamma
q^2\right) \label{lubricationdispersion}
\end{equation}
where $q$ is the wave number of the perturbation.

In order to determine the shape of the moving dewetting front, we
now look for travelling-wave solutions of the form
$\zeta(x,t)=\zeta(x-vt)$, where $v$ is the velocity of the
dewetting front. We remain in Fourier space, i.e., we decompose
the front profile into modes $\zeta_q \propto \exp(iqx-\omega t)$.
Writing $Q:=iqh$, it follows that $\omega = Qv/h$, and from
eq.~(\ref{lubricationdispersion}) we then get
\begin{equation}
Q^4 - \frac{\alpha E h}{\gamma } Q^2 - \frac{v h^3}{\gamma C} Q +
\frac{\omega_0  h^4}{\gamma C} = 0.
\label{Quartic}
\end{equation}
As it will become clear below, the term $\omega_0 h^4 / \gamma C$
is rather small. If we neglect it for the moment, we get the cubic
equation
\begin{equation}
Q^3 - \tilde{E} Q - \tilde{v} = 0 \label{Cubic}
\end{equation}
where we have introduced the dimensionless quantities
$\tilde{v}:=\frac{v h^3}{C\gamma}$ and $\tilde{E}:=\frac{\alpha E
h}{\gamma}$. This can be easily solved,  and the solutions of the
quartic equation~(\ref{Quartic}) turn out to deviate significantly
from those of the cubic equation only for parameters which are not
relevant in most experimental situations. We will thus primarily
discuss eq.~(\ref{Cubic}), and refer to eq.~(\ref{Quartic}) only
occasionally, where appropriate.

If viscoelastic effects are absent, i.e. $E = 0$, we have $Q =
\sqrt[3]{\tilde{v}}$. We are interested only in modes with
$\Re(Q)< 0$ \cite{RealImag}, which are $Q =
-\frac{1}{2}\sqrt[3]{\tilde{v}}(1 \pm i\sqrt{3})$. The shape of
the real surface of the leading front is thus $\propto
e^{-x}\cos(\sqrt{3}x)$ \cite{Unterschwinger,Srolovitz2}, which
clearly exhibits not only a ditch, but a damped oscillation. For
Newtonian liquids, it can be seen from numerical solutions of the
Navier Stokes equation for the moving front problem that this
result does not depend on the friction of the film at the
substrate \cite{Spaid,PhysicsOfFluids}. A profile like that is
shown in fig.~\ref{Observations}b, and it is obtained only for
small molecular weight indeed, where elastic effects are
particularly small. The real and imaginary part of $Q$ can be
determined from the profile, and in the case of the one displayed
in fig.~\ref{Observations}b, it turns out that they are not
precisely as expected for a Newtonian fluid \cite{Unterschwinger}.
Since the complex roots of eq.~(\ref{Cubic}) satisfy the relation
\begin{equation}
3\Re(Q)^2-\Im(Q)^2=\tilde{E}
\end{equation}
the elasticity of the film can be inferred from the measurements,
and we get $\alpha E = $ 8.7 kPa. We would like to point out that
this is a non-invasive method of determining rheological
properties of liquids on a sub-micron scale.

As the height of the crest, $H$, increases while more material is
collected in the front, the amplitude of the oscillation
increases, and so does the depth of the ditch
\cite{Unterschwinger}. When the latter reaches the substrate, the
crest is pinched off and a cascade is formed as shown in
fig.~\ref{Observations}c. However, this works only as long as the
width, $W$, of the crest is less than half the wavelength of the
damped oscillation. As $H$ increases, so does $W$
\cite{LangmuirRupture,BrochardExtrapLength}, with the ratio $H/W
=: G(\Theta)$ being only a function of the dynamic contact angle
at the substrate, $\Theta$ \cite{LangmuirRupture}. For a
cylindrical rim \cite{BrochardExtrapLength}, we have simply
$G(\Theta)=(1-\cos\Theta)/2\sin\Theta$. For more asymmetric
shapes, the form of $G$ is different, but similar as to the
overall behavior and the order of magnitude. If $W$ is larger than
$W_c = \lambda/2 = \pi/\Im(Q)$, the depth of the ditch is rather
determined by the contact angle of the rim at the leading edge.
Consequently, it will reach a final maximal depth
\cite{Unterschwinger}, and a dry spot forms only if this depth
exceeds the film thickness. The condition for a cascading front to
form is readily seen to be
\begin{equation}
\pi G(\Theta) > \mid\Im(Q)\mid \cdot
\exp\left(\pi\frac{\mid\Re(Q)\mid}{\mid\Im(Q)\mid}\right)
\label{finaldepth}
\end{equation}
As the film thickness is decreased, so is the right hand side of
eq.~(\ref{finaldepth})($h$ cancels out in the exponential), such
that a cascade is expected at sufficiently small $h$.
Eq.~(\ref{finaldepth}) can be used to determine the boundary line
for the appearance of cascade structures in the
$(\tilde{v},\tilde{E})$-plane. In fig.~\ref{PhaseDiagram}a, this
is indicated by the lower curve for $G = \pi^{-1}$. For different
values of $G$,  the maximum of the boundary line, indicated by the
dot in the figure, is quite accurately described by $\tilde{E} =
0.060 \cdot G^2$. For very thin films, where van der Waals forces
play a role, the cascade region will slightly extend or shrink,
depending on the sign of the Hamaker constant.

Let us now turn to the influence of stronger elastic effects. It
is interesting to investigate at which system parameters the
damped oscillation, as shown in fig.~\ref{Observations}b,
vanishes. These are given by the zero of the discriminant of the
cubic equation~(\ref{Cubic}), which is
\begin{equation}
D = \left(\frac{\tilde{v}}{2 }\right)^2 - \left(\frac{\tilde{E}}{3
}\right)^3 \label{Diskriminante}
\end{equation}
The critical modulus above which there is no oscillation is thus
given by $\tilde{E}=3\left(\frac{\tilde{v}}{2}\right)^{2/3}$. This
may be viewed as another morphological `phase boundary' in
the($\tilde{E}$,$\tilde{v}$)-plane, and is shown in
fig.~\ref{PhaseDiagram}a as the upper solid curve. The exact
boundary, according to  eq.~(\ref{Quartic}), deviates noticeably
only for very small values of $\tilde{E}$ and $\tilde{v}$. To
quantify this deviation, it is useful to introduce the
dimensionless parameter $\Omega:= \omega_0 h/v$, which represents
the importance of intrinsic relaxation in the material with
respect to the shear flow exerted by the dewetting process itself.
In fig.~\ref{PhaseDiagram}b, the `exact' phase boundary, which was
determied by solving eq.~(\ref{Quartic}) graphically, is shown as
the solid curve, along with the approximation obtained from the
cubic equation (dashed line). For $\Omega \rightarrow 0$, the
solid curve approaches the dashed line asymptotically. The diagram
presented in fig.~\ref{PhaseDiagram}a, which corresponds to
$\Omega = 0$, shall thus be considered quite reliable  within the
range displayed if $\Omega < 0.1$ (in the experiments in
ref.~\cite{falaises}, $\Omega \approx 10^{-5}$).

The two complex conjugate roots giving rise to the oscillations
for $D > 0 $ merge into a degenerate real root at a critical
elastic modulus ($D=0$). As the latter increases further, they
bifurcate again into two real roots, $Q_{long}$ and $Q_{short}$.
In fact, `elastic' fronts can be well fitted by a superposition of
two exponentials, as shown in fig.~\ref{Observations}a by the
dotted curves. The model predicts the ratio $Q_{short}/Q_{long}$
to increase with $\tilde{E}$, and thus the front to become more
and more asymmetric, in accordance with observation
\cite{Karinsdiss,falaises,Unterschwinger}. We can again determine
$\tilde{E}$ from the front profile via the relation
\begin{equation}
\frac{Q_{long}}{Q_{short}} = \frac{\sin
\{\frac{\pi}{6}-\frac{1}{3}\arccos B \}}{\sin
\{\frac{\pi}{6}+\frac{1}{3}\arccos B \}}
\end{equation}
with $B:=\frac{3}{2}\sqrt{3}\tilde{v}\tilde{E}^{-3/2}$, once the
constant $C$ is known for the system.

The long range shape of the front is determined by the smaller
root, $Q_{long}$, which asymptotically approaches the simple law
$Q = -\tilde{v}/\tilde{E}$. Thus the width of an elastic front
should remain constant during dewetting, in agreement with recent
observations \cite{falaises}, but in contrast to quasi-Newtonian
fronts \cite{LangmuirGrowth,BrochardExtrapLength}. Since our model
predicts $Q_{short}$ to scale as $\sqrt{\tilde{E}}$ for
sufficiently large $\tilde{E}$, we have $Q_{long} \propto
Q_{short}^2$ if all other parameters in the system are kept
constant. In fact, for the two curves shown in
fig.~\ref{Observations}a, the ratio of the $Q_{long}$ is $1.40 \pm
0.07$, while the ratio of the $Q_{short}^2$ is $1.34 \pm 0.13$.
This is clear agreement.

We thus have put forward a tractable model which describes the
generic morphologies of liquid dewetting fronts observed so far.
It correctly accounts for the impact of the fluid properties upon
their occurrence and structure. It is thereby capable of
extracting information on the sub-micron scale rheological
properties of the liquid film from the observed profiles. This may
be rendered quantitative by determining the numerical constant
$\alpha$, which we leave to further work.

The authors are indebted to Klaus Mecke and Ralf  Blossey for many
fruitful discussion. We also thank G\"unter Reiter for stimulating
remarks. Funding from the Deutsche Forschungsgemeinschaft within
the Priority Program `Wetting and Structure Formation at
Interfaces' is gratefully acknowledged.

\begin{figure}
\caption{Dewetting morphologies observed in various liquid
polystyrene films beading off a silicon substrate. The height has
been normalized with respect to the film thickness. a) Profiles of
the rims of holes nucleated in 40 nm thick films. The steep
profile was obtained with a molecular weight of ${\rm M_W = }$ 101
k, the flatter profile with ${\rm M_W = }$ 600 k. Each profile is
superimposed with a double exponential fit as a dotted line, the
agreement is almost perfect (see text). b) Profile obtained with
${\rm M_W = }$ 2 k. c) Same as b), but with smaller film
thickness.} \label{Observations}
\end{figure}

\begin{figure}
\caption{a) Morphological phase diagram for viscoelastic dewetting
fronts in the $(\tilde{E},\tilde{v})$-plane, according to
eq.~(\protect\ref{Cubic}). b) Deviation of the phase boundary
according to eq.~(\protect\ref{Quartic}) (solid curve) from the
approximation given by eq.~(\protect\ref{Cubic}) (dashed line).}
\label{PhaseDiagram}
\end{figure}


\begin{references}

\bibitem{Srolovitz1}D. J. Srolovitz, S. A. Safran, {\it J. Appl. Phys.}
{\bf 60} (1986) 247.

\bibitem{Redon}C. Redon, F. Brochard-Wyart, F. Rondelez, {\it Phys. Rev.
Lett.} {\bf 66} (1991) 715.

\bibitem{Reiter} G. Reiter, {\it Phys. Rev. Lett.} {\bf 68} (1992)
75.

\bibitem{Karinsdiss} K. Jacobs, {\it Thesis}, Konstanz 1996;
ISBN 3-930803-10-0.

\bibitem{Lambooy} P. Lambooy, K. C. Phelan, O. Haugg, G. Krausch,
{\it Phys. Rev. Lett.} {\bf 76} (1996) 1110.

\bibitem{Stange} T. G. Stange, D. F. Evans, W. A. Hendrickson,
{\it Langmuir} {\bf 13} (1997) 4459.

\bibitem{Science} S. Herminghaus et al., {\it Science} {\bf 282}
(1998) 916.

\bibitem{Moriarty} J. Moriarty, L. Schwartz, E. Tuck, {\it Phys. Fluids A}
{\bf 3} (1991) 733.

\bibitem{Greenspan} H. Greenspan, {\it J. Fluid Mech.} {\bf 84}
(1978)125.

\bibitem{Troian} S. Troian, E. Herbolzheimer, S. Safran, J. F.
Joanny, {\it Europhys. Lett.} {\bf 10} (1989) 25.

\bibitem{Spaid} M. Spaid, G. Homsy, {\it J. Non-Newt. Fluid Mech.}
{\bf 55} (1994) 249.

\bibitem{PhysicsOfFluids} M. A. Spaid, G. M. Homsy, {\it Phys. Fluids} {\bf
8} (1996) 460.

\bibitem{falaises} G. Reiter, {\it Phys. Rev. Lett.} {\bf 87}
(2001) 186101.

\bibitem{Unterschwinger} R. Seemann, S. Herminghaus, K. Jacobs,
{\it Phys. Rev. Lett.} {\bf 87} (2001) 196101.

\bibitem{LangmuirGrowth} K. Jacobs, R. Seemann, G. Schatz, S.
Herminghaus, {\it Langmuir} {\bf 14} (1998) 4961.

\bibitem{GuenterMacromolecules} G. Reiter, P. Auroy, L. Auvray, {\it Macromolecules}
{\bf 29} (1996) 2150.

\bibitem{LangmuirRupture} K. Jacobs, S. Herminghaus, K. Mecke, {\it Langmuir}
{\bf 14}(1998) 965.

\bibitem{Nucleation} D. Podzimek et al., cond-mat/0105065 (2001).

\bibitem{LL1} L. D. Landau, E. M. Lifshitz, {\it Theory of Elasticity} (Vol.
{\bf VII}), Butterworth, London 1995.

\bibitem{LL2} L. Landau, Lifshitz, {\it Hydrodynamics} (Vol. {\bf VI}),
Butterworth, London 1995.

\bibitem{Questions} S. Herminghaus, K. Jacobs, R. Seemann, {\it Eur. Phys. J. E}
{\bf 5} (2001) 531.

\bibitem{SurfaceMelting} S. Herminghaus, {\it Eur. Phys. J. E}
(2002)in print.

\bibitem{BrochardExtrapLength} F. Brochard-Wyart, P.-G. de Gennes,
H. Hervert, C. Redon, {\it Langmuir} {\bf 10} (1994) 1566.

\bibitem{Reconciliation} R. Seemann, S. Herminghaus, K. Jacobs,
{\it Phys. Rev. Lett.} {\bf 86} (2001) 5534.

\bibitem{RealImag} $\Re(\cdot)$ and $\Im(\cdot)$ denote the real and imaginary
part of their argument, respectively.

\bibitem{Srolovitz2} The same result is obtained for surface diffusion
kinetics (D. J. Srolovitz, S. A. Safran, {\it J. Appl. Phys.}
{\bf 60} (1986) 255).

\end{references}
\end{document}